\documentclass{emulateapj}
\shorttitle{The Third Transit of \textit{Kepler}-421b}
\shortauthors{Dalba \& Muirhead}

\usepackage{subfigure}
\usepackage{url}
\usepackage{hyperref}
\usepackage{epsfig}
\usepackage{graphicx}
\usepackage{sidecap}
\usepackage{hanging}
\usepackage{color}
\usepackage{multirow}
\usepackage{enumerate}
\usepackage{natbib}
\usepackage{amssymb}
\usepackage{pdflscape}
\usepackage{amsmath}
\usepackage{enumitem}
\setlist[description]{leftmargin=\parindent,labelindent=\parindent}

\newcommand\icarus{Icarus }

\newcommand{\Kepler}{{\it Kepler} }

\newcommand{\Spitzer}{{\it Spitzer} }

\providecommand{\e}[1]{\ensuremath{\times 10^{#1}}}

\begin{document}

\title{No Timing Variations Observed in Third Transit of Snow-Line Exoplanet \textit{Kepler}-421b}

\author{Paul A. Dalba\altaffilmark{1,2} and Philip S. Muirhead\altaffilmark{1,2}}

\affil{\vspace{0pt}\\ $^{1}$Department of Astronomy, Boston University, 725 Commonwealth Ave., Boston, MA, 02215, USA; \href{mailto:pdalba@bu.edu}{pdalba@bu.edu} \\
$^{2}$Institute for Astrophysical Research, Boston University, 725 Commonwealth Ave., Boston, MA, 02215 USA}

\begin{abstract}

We observed \textit{Kepler}-421 during the anticipated third transit of the snow-line exoplanet \textit{Kepler}-421b in order to constrain the existence and extent of transit timing variations (TTVs). Previously, the \Kepler Spacecraft only observed two transits of \textit{Kepler}-421b leaving the planet's transit ephemeris unconstrained. Our visible light, time-series observations from the 4.3-meter Discovery Channel Telescope were designed to capture pre-transit baseline and the partial transit of \textit{Kepler}-421b barring significant TTVs. We use the light curves to assess the probabilities of various transit models using both the posterior odds ratio and the Bayesian Information Criterion (BIC) and find that a transit model with no TTVs is favored to 3.6$\sigma$ confidence. These observations suggest that \textit{Kepler}-421b is either alone in its system or is only experiencing minor dynamic interactions with an unseen companion. With the \textit{Kepler}-421b ephemeris constrained, we calculate future transit times and discuss the opportunity to characterize the atmosphere of this cold, long-period exoplanet via transmission spectroscopy. Our investigation emphasizes the difficulties associated with observing long-period exoplanet transits and the consequences that arise from failing to refine transit ephemerides.  
\end{abstract}

\maketitle

\section{Introduction}

Since the end of the primary \Kepler Mission, the number of confirmed or validated transiting exoplanets with periods of hundreds of days has grown rapidly, despite their low inherent transit probabilities. Presently, the NASA Exoplanet Archive reports 138 confirmed transiting exoplanets with periods greater than 100 days, 9 of which have periods greater than 500 days \citep{Akeson2013}. This is in part due to dedicated searches for long-period exoplanets \citep[e.g.][]{Fischer2012,Wang2015,Uehara2016} and is also a result of the ongoing analysis of the \Kepler data set \citep[e.g.][]{Morton2016,Kipping2016b}. 

Long-period exoplanets that orbit near or beyond their systems' snow-lines provide valuable opportunities to explore theories of planetary formation and evolution \citep{Oberg2011}. Their atmospheric abundances provide a test of theories such as core accretion \citep{Pollack1996} and disk instabilities \citep{Boss1997}. Furthermore, massive long-period planets may influence the dynamic stability and final architecture of multi-planet systems through planetary migration \citep{Rasio1996,Ida2004}. 

A variety of observational challenges have so far impeded efforts to probe the atmospheres of exoplanets akin to the four gas giants in the solar system. Their low effective temperatures prevent detection through direct imaging. These exoplanets' slow orbital velocities also limit the usefulness of high-resolution spectroscopy to observe Doppler-shifted, atmospheric spectral features \citep{Snellen2010,Birkby2013}. Characterizing long-period exoplanet atmospheres through reflected light is possible but will not be feasible until the launch of future missions such as the Wide-Field Infrared Survey Telescope \citep{Spergel2015,Lupu2016}. 

The atmospheres of long-period transiting exoplanets can be probed via transmission spectroscopy \citep{Dalba2015}. However, for an exoplanet with fewer than three documented transits, the magnitude of transit timing variations (TTVs) due to the dynamical influence of an unknown companion is unconstrained. In some cases, long-period exoplanets exhibit TTVs of up to 40 hours \citep{Wang2015}. Without an estimate of the magnitude of TTVs, planning follow-up observations for atmospheric characterization is a risky practice, if not altogether infeasible. 

Here we focus on the long-period exoplanet \textit{Kepler}-421b: a 4.2-$R_{\earth}$ planet transiting a G9V star with an orbital period of $\sim$704 days discovered by \citet{Kipping2014a}. \textit{Kepler}-421b has a calculated equilibrium temperature of $\sim$185 K (assuming a Bond albedo of 0.3) and likely resides near its system's snow-line. The \textit{Kepler} Spacecraft only observed two 15.8-hour transits of \textit{Kepler}-421b in Quarters 5 and 13. Since the re-purposed \textit{K2} Mission cannot return to the original \Kepler field \citep{Howell2014}, there was no dedicated observatory to view the third transit that was set to occur on UT 2016 February 19. With only two documented transits and no information about companions in the \textit{Kepler}-421 system, the influence of TTVs on \textit{Kepler}-421b's transit ephemeris was entirely unknown.   

In this letter, we describe ground-based observations of \textit{Kepler}-421 taken during the anticipated third transit of \textit{Kepler}-421b (assuming a linear ephemeris) in order to constrain the magnitude of TTVs.

\section{Observations}\label{sec:obs}

We observed \textit{Kepler}-421\footnote{KOI-1274, KIC-8800954, $\alpha=283\fdg256810$, $\delta=+45\fdg087780$} in V-band on the calendar mornings of 2016 February 18, 19, and 20 with the Large Monolithic Imager \citep[LMI;][]{Massey2013} on the 4.3-meter Discovery Channel Telescope (DCT). Each observation began at approximately 01:00 Mountain Standard Time (MST) and concluded in civil twilight ($\sim$06:45 MST). 

In the absence of TTVs and assuming a linear ephemeris, the Barycentric Kepler Julian Date (BKJD)\footnote{BKJD is defined by BJD = BKJD + 2454833.0 where BJD is the Barycentric Julian Date.} of mid-transit ($t_0$) was 2605.3626$\pm$0.0030 or approximately 13:42 MST on 2016 February 19. We will hereafter refer to this as the \emph{nominal} transit of \textit{Kepler}-421b. Using the transit duration measured from the \Kepler data \citep[15.79$^{+0.12}_{-0.10}$ hours,][]{Kipping2014a}, the approximate start of ingress was 05:48 MST with an uncertainty of 8.4 minutes. The airmass of \textit{Kepler}-421 at the nominal ingress was 1.34 and rising. Astrometric twilight began at 05:44 MST followed by nautical and civil twilight at 06:13 and 06:43 MST, respectively. 

The exposure times were 10, 15, and 3 seconds for the 3 nights, respectively. These times maximized the amount of signal in each integration while still staying safely below the LMI's nonlinear response regime. The weather during the first two nights consisted of partly cloudy skies and occasional cirrus layers.  Periods of increased cloudiness during the night of the nominal transit are apparent in the light curve of \textit{Kepler}-421. Conditions during the third night were mostly clear. 

The analysis described in the following sections was only applied to exposures taken while \textit{Kepler}-421 was at a relatively low airmass, ranging from 2.0--1.2.

\section{Data Analysis}\label{sec:analysis}

\subsection{Calibration}

We developed a custom pipeline for the calibration and analysis of time-series, photometric observations taken with the DCT-LMI. The calibration consisted of bias and over scan subtraction and a flat-field correction. 

The dark current on the LMI was estimated to be 0.07 electrons pixel$^{-1}$ hour$^{-1}$, which was negligible for these observations.

\subsection{Differential Aperture Photometry}\label{sec:aper_phot}

We conducted differential aperture photometry on \textit{Kepler}-421 using calibration stars in the 12$\farcm$3x12$\farcm$3 field of view of the DCT-LMI. Our initial source extraction identified several hundred sources in the field, but we only conducted photometry on a subset of these (hereafter referred to as ``calibration stars'') that were not saturated or located too close to another star or the edge of the detector. 

For each image, we conducted aperture photometry for \textit{Kepler}-421 and each calibration star using apertures of radius 21, 15.5, and 12.5 pixels for each night, respectively. These apertures produced the lowest out-of-transit scatter in the final light curves. In every case, we estimated and removed the local background signal for each source using sky annuli with inner and outer radii of 25 and 35 pixels. The resulting light curve of \textit{Kepler}-421 and each calibration star was then normalized to its own median value.

\begin{figure*}
\centering
\includegraphics[scale=0.5]{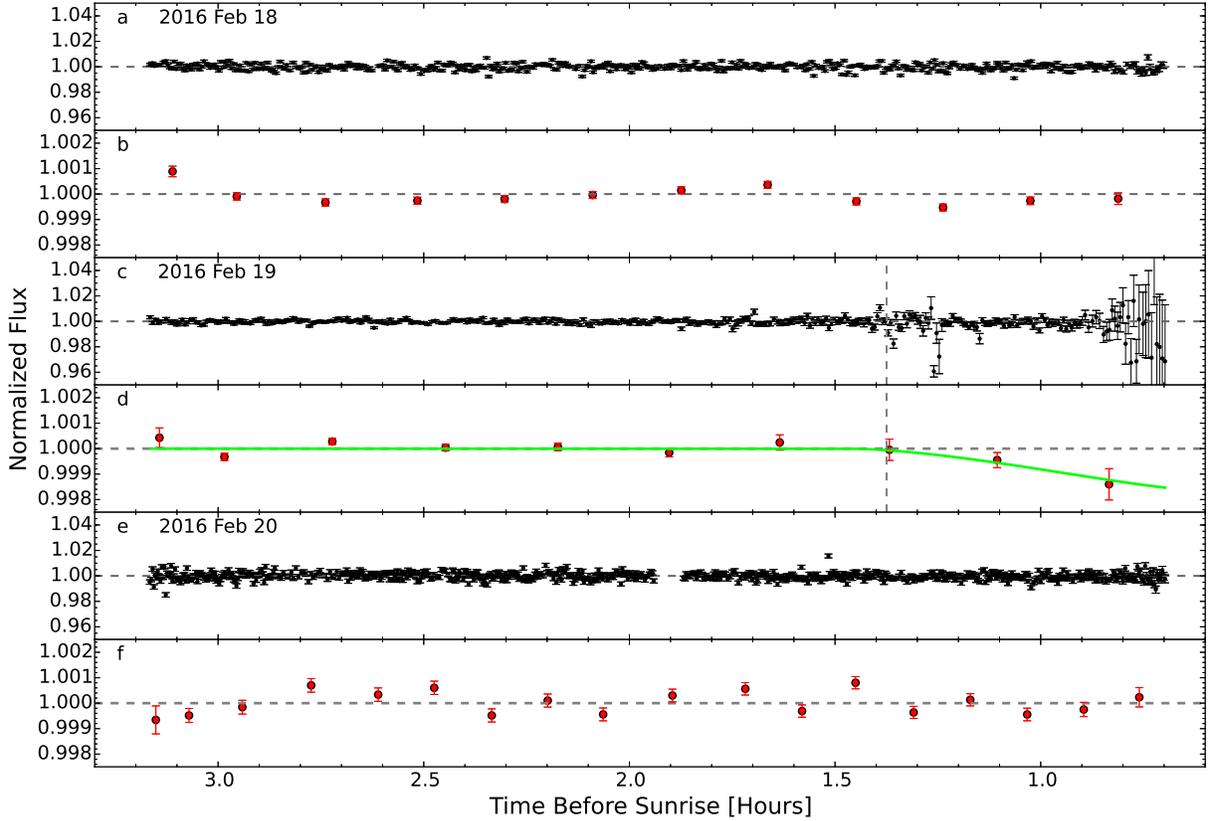}
\caption{DCT-LMI photometry of \textit{Kepler}-421 from the mornings of 2016 February 18 (panels a and b), 2016 February 19 (panels c and d), and 2016 February 20 (panels e and f). The black data points are individual exposures and the red data points are variance-weighted and binned by a factor of 40. Only data taken at an airmass less than two are shown. There were 460, 368, and 690 data points for 2016 February 18, 19, and 20, respectively. The horizontal dashed lines are drawn at unity for reference. The vertical dashed lines in panels c and d are drawn at the nominal ingress time of the third transit of \textit{Kepler}-421b. The green line in panel d is a theoretical model of the nominal transit of \textit{Kepler}-421b and has not been binned or fit to the data.}
\label{fig:lc_all}
\end{figure*}

We further normalized the light curve of \textit{Kepler}-421 to that of a \emph{master} calibration star, which was the mean light curve of a subset of all of the calibration stars in the image. To determine this subset, we first identified the 15 calibration stars whose light curves exhibited the smallest residuals compared to that of \textit{Kepler}-421. Then we calculated the final normalized light curve of \textit{Kepler}-421 using every combination of those 15 calibration sources and chose the set that yielded the lowest out-of-transit scatter. For consistency, the set of calibration sources selected via this procedure (Table \ref{tab:cal_stars}) for 2016 February 19 was also employed for the other two nights. 

\begin{deluxetable}{cc}
\tablecaption{Final Calibration Stars\label{tab:cal_stars}} 
\tablehead{ 
\colhead{KIC ID\tablenotemark{a}} & \colhead{\Kepler Magnitude} }
\startdata 
8800901  & 14.789  \\
8866248  & 14.483  \\
8866315  & 14.437  \\
8800997  & 14.263  \\
8801075  & 14.387  \\
8800921  & 13.699  
\enddata 
\tablenotetext{a}{\Kepler Input Catalog \citep{Brown2011}.}
\end{deluxetable}

A slight background trend was present in each light curve. Assuming that this signal was a linear function of time, we fit and removed the trend via a least squares linear regression of the data collected before 05:15 MST each night ($\sim$0.5 hours prior to the nominal time of ingress on 2016 February 19). 

The final light curves of \textit{Kepler}-421 for each night are shown in Fig. \ref{fig:lc_all}. The errors on the individual exposures included the Poisson noise from the star and background and the read noise, which was $\sim$6 electrons per pixel. These uncertainties were propagated through the calculation of the master calibration star and the normalization of the \textit{Kepler}-421 light curves. Also shown in Fig. \ref{fig:lc_all} are light curves of \textit{Kepler}-421 that have been binned by a factor of 40. 

Prior to three hours before sunrise, \textit{Kepler}-421 had an airmass near 2 and the data points displayed a minor increase in signal. This was likely a result of slight color differences between the target and calibration stars integrated over the long columns of air. The effect was not apparent in the third night when the integration time was much shorter. 

The light curve of \textit{Kepler}-421 from 2016 February 19 exhibited a slight decrement in flux coinciding with the transit ingress of exoplanet \textit{Kepler}-421b in the absence of large TTVs. The feature was not immediately obvious from the individual exposures but became clear when the signal was binned. Similar features were not present in the photometry from the other two nights of observation. 

\begin{figure}
\centering
\includegraphics[scale=0.45]{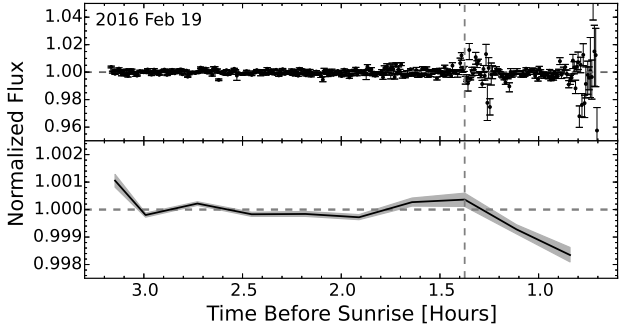}
\caption{Distribution of \textit{Kepler}-421 light curves created from 1000 sets of calibration stars for 2016 February 19. In the top panel, the black data points represent the 50th percentile of the distribution and their error bars represent the 16th and 84th percentiles. In the bottom panel, the points have been binned by a factor of 40 and the gray region represents the 1$\sigma$ spread in the binned points. The vertical dashed lines are drawn at the nominal ingress time of \textit{Kepler}-421b. The decrease in flux near the end of the observation persists even when other sets of calibration stars are used to normalize the \textit{Kepler}-421 photometry.}
\label{fig:ref_var}
\end{figure}

To test if the feature was introduced by one of the calibration stars, we recalculated the light curve of \textit{Kepler}-421 1000 times using different sets of calibrations stars. These sets were the combinations of the original 15 calibration stars mentioned previously that yielded the lowest out-of-transit scatter in the final light curve. The resultant distribution of light curves is presented in Fig. \ref{fig:ref_var} and the decrease in flux at the nominal time of ingress is still present. The excess of flux caused by the increasing airmass prior to three hours before sunrise is also present.

We also visually inspected the normalized light curves of each calibration star used in the analysis and failed to find any spurious features.  

\section{Hypothesis Testing}\label{sec:hyp_test}

The moderate precision achieved by the photometry and the subtlety of the transit signature of \textit{Kepler}-421b (transit depth $\approx$ 0.3\%) warranted a careful analysis of the \textit{Kepler}-421 light curves. We considered three simple hypotheses regarding the third transit of \textit{Kepler}-421b\footnote{All theoretical transit curves were simulated using the analytical relations of \citet{Mandel2002}.}:

\begin{description}
\item[$H_0$:] The transit of \textit{Kepler}-421b did not display TTVs and began on 2016 February 19 according to a linear ephemeris. The light curves from the other two nights of observation displayed no variation. The transit parameters for this hypothesis were taken from \citet{Kipping2014a} and the mid-transit time ($t_0$) was simply extrapolated linearly from the previous two transits. This hypothesis had zero free parameters. 
\item[$H_1$:] The final light curves of \textit{Kepler}-421 displayed no variation. This comprehensive hypothesis included scenarios where the transit occurred before the observations on 2016 February 18, after the observations on 2016 February 20, or during a gap in the observations. It may also suggest that the precision of the photometry was too poor to distinguish the transit from a flat line. This hypothesis had zero free parameters.
\item[$H_2$:] The transit of \textit{Kepler}-421b occurred with a TTV such that the observations captured some portion of the transit. We considered TTVs in the ranges [$-1.74$,$-0.97$], [$-0.75$,$0.03$], and [$0.26$,$1.03$] days, which shifted transit egress just before the first observation on 2016 February 18 or transit ingress just after the final observation on 2016 February 20. Away from transit, the light curves displayed no variation. All transit parameters with the exception of $t_0$ were taken from \citet{Kipping2014a}. This hypothesis had one free parameter: $t_0$. 
\end{description}

\subsection{Odds Ratio}

We assessed the plausibility of our hypotheses by calculating the posterior odds ratio 

\begin{equation}
O_{ij} = \mathcal{P}_{ij} \; B_{ij} 
\end{equation}

\noindent where  $\mathcal{P}_{ij}$ was the prior odds and $B_{ij}$ was the ratio of Bayesian evidences, also known as the Bayes factor \citep[see, for instance,][]{Gregory2005}, for hypotheses $i$ and $j$. Without prior information, we assumed $\mathcal{P}_{ij}$ = 1 for the comparison of each set of hypotheses. We also assumed Gaussian likelihood functions when calculating the Bayesian evidences. 

For $H_0$ and $H_1$, which had zero free parameters, $B_{01}$ was simply the likelihood ratio 

\begin{equation}
B_{01} = \exp{\left [-\frac{1}{2} (\chi^2_0 - \chi^2_1) \right ] }
\end{equation}

\noindent where $\chi^2_0$ and $\chi^2_1$ were the chi-squared statistics for $H_0$ and $H_1$, respectively. These two hypotheses only differed during the final $\sim$0.7 hours of observation on 2016 February 19. For the unbinned \textit{Kepler}-421 light curves, $B_{01} = $ 134, offering \emph{moderate} to \emph{strong} evidence in favor of $H_0$ according to the Jeffreys' scale \citep{Jeffreys1961,Kass1995}. To aid in the interpretation of this result, we approximated the frequentist $p$-value test statistic ($p$) using $B_{ij} \leq -(e \, p \, \ln{p})^{-1}$ where $e$ is the base of the natural logarithm \citep{Sellke2001}. For $B_{01} = $ 134, $p = $ 3.43$\e{-4}$ meaning that $H_0$ was favored to 3.6$\sigma$ confidence.

$H_2$ was inherently more complicated than $H_0$ and $H_1$ because it included the free parameter $t_0$, the mid-transit time. The calculation of $B_{02}$, the Bayes factor between $H_0$ and $H_2$, necessitated a prior on $t_0$. Since the potential for TTVs in the \textit{Kepler}-421 system was unknown, we used a uniform prior over the range of $t_0$ such that the observations captured some portion of the transit. The lengths of these ranges were 0.77, 0.78, and 0.77 days for the three nights of observation, respectively. We then integrated the likelihood function over these ranges and computed the Bayesian evidence for $H_2$, which we divided into the likelihood of $H_0$. This yielded the Bayes factor $B_{02} = 370$, meaning that $H_0$ was favored to 3.9$\sigma$.

\subsection{Bayesian Information Criterion}

To verify the result suggested by the Bayes factors, we also assessed the plausibility of our hypotheses with the criterion of \citet{Schwarz1978}, presented as the Bayesian information criterion (BIC):

\begin{equation}\label{eq:bic}
BIC_i = -2 \ln{M_i} + k_i \ln{n}
\end{equation}

\noindent where $M_i$ was the maximum of the likelihood function, $k_i$ was the number of free parameters, and $n$ was the number of data points for each hypothesis $i$. The BIC did not require the specification of priors, and the most plausible hypothesis was that which minimized the BIC.

By employing the BIC in our analysis, we assumed that the data were independent and identically distributed. As in our calculations of the Bayes factors, we once again assumed Gaussian likelihood functions. 

We determined the BIC values for the three hypotheses using the unbinned photometry of \textit{Kepler}-421. Since both $H_0$ and $H_1$ had zero free parameters, the comparison of their BIC values ($\Delta$BIC$_{H_1}$ = BIC$_{H_1}$ $-$ BIC$_{H_0}$ = +9.8) was identical to the calculation of $B_{01}$ presented above.

\begin{figure}
\centering
\includegraphics[scale=0.45]{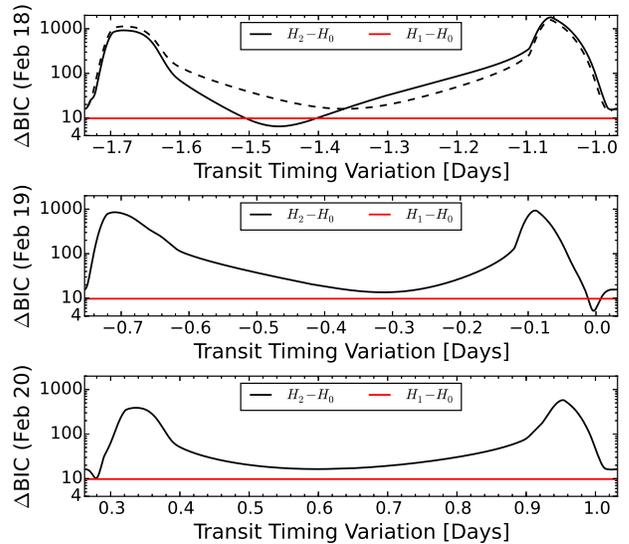}
\caption{Differences in BIC values between $H_2$ (a \textit{Kepler}-421b transit with TTVs) and $H_0$ (the nominal \textit{Kepler}-421b transit without TTVs) calculated from the unbinned data. The red line at $\Delta$BIC$=$+9.8 shows the difference in BIC between $H_1$ (no variability in the light curves) and $H_0$. This value does not depend on TTV length but is drawn for reference to the black lines. The dashed black line in the top panel shows the dependence of the $\Delta$BIC values for 2016 February 17 on the background removal procedure. Since all $\Delta$BIC values are positive with respect to the nominal transit hypothesis ($H_0$), the most likely explanation of the data set is that the transit of \textit{Kepler}-421b occurred without TTVs according to a linear ephemeris.}
\label{fig:ttv_bic}
\end{figure}

For $H_2$, we calculated BIC values for TTVs in the previously mentioned ranges sampled every 0.5 minutes.  The differences in BIC values relative to $H_0$ ($\Delta$BIC$_{H_2}$ = BIC$_{H_2}$ $-$ BIC$_{H_0}$) are shown in Fig. \ref{fig:ttv_bic} along with $\Delta$BIC$_{H_1}$ for reference. For all TTVs considered, $\Delta$BIC$_{H_2}$ was positive, which offered further evidence that the transit of \textit{Kepler}-421b occurred without TTVs. 

The difference in BIC values between $H_2$ and $H_1$ was also positive, suggesting that $H_1$ was more plausible that $H_2$ given the data. This conclusion was especially valid for the TTV values near the six peaks in the $\Delta$BIC distributions that corresponded to TTVs placing transit ingress or egress near the middle of the \textit{Kepler}-421 light curves. These scenarios were clearly ruled out by the DCT-LMI photometry. The troughs in between the $\Delta$BIC peaks corresponded to TTVs that placed the observations in mid-transit of \textit{Kepler}-421b, which closely resembled a flat line (i.e. $H_1$). For these TTVs, the $\Delta$BIC was primarily a result of the ``Occam's Razor'' penalty for the added free parameter ($t_0$). 

In two cases, the BIC$_{H_2}$ dipped below BIC$_{H_1}$ despite the $k \ln{n}$ penalty. First, near a TTV value of -1.46 days, the seemingly flat light curve of \textit{Kepler}-421 appeared to resemble a portion of the theoretical transit of \textit{Kepler}-421b before mid-transit. However, if the slight background trend present in the data was removed using the entire night's observations versus a subset matching 2016 February 19 (see \S\ref{sec:aper_phot}), the feature vanished and the entire $\Delta$BIC curve closely resembled the other nights' curves (Fig. \ref{fig:ttv_bic}, dashed line). A recalculation of the posterior odds ratio between $H_0$ and $H_2$ using this alternative background subtraction procedure yielded $B_{02} = 1283$, or 4.2$\sigma$, strengthening the conclusion the the data favored the hypothesis $H_0$.

The second case of BIC$_{H_2}$ $<$ BIC$_{H_1}$ occurred near a TTV of zero days. As opposed to the first case, this feature was not sensitive to the background removal procedure. The minimum occurred at a TTV of -5.0 minutes where BIC$_{H_2}$ $-$ BIC$_{H_1}$ = -4.6. Considered independently, this provided \emph{moderate} evidence in favor of $H_2$, with a TTV of 5 minutes, over $H_1$. However, considering the full width at half maximum of the $\Delta$BIC feature ($\sim$20 minutes) and the uncertainty in the time of ingress (8.4 minutes, \S\ref{sec:obs}), we instead interpreted this result as further support of the hypothesis that the third transit of \textit{Kepler}-421b occurred without measurable TTVs.

\section{Discussion}\label{sec:discussion}

Based on observations taken with the DCT-LMI, the third transit of \textit{Kepler}-421b likely occurred without TTVs. Either \textit{Kepler}-421b is the only planet in its system or the dynamical interactions with undiscovered companions are too weak to alter \textit{Kepler}-421b's orbit. Without significant TTVs, the future transits of \textit{Kepler}-421b may be predicted by linearly extrapolating its transit ephemeris. In Table \ref{tab:transits}, we predict the next six transits and propagate errors based on uncertainties in the period and $t_0$ of the first two transits, and the uncertainty in ingress time of the third transit ($\sim$20 minutes, \S\ref{sec:hyp_test}). 

The next transit is due to occur at UT 01:27 on 2018 January 24 with a margin of error of $\sim$40 minutes. With ingress and egress $\sim$8 hours before and after this date, this transit will not be observable from the ground. The fortuitous circumstance described in this letter whereby out-of-transit baseline and a portion of transit were both observed at reasonable airmass from a ground-based observatory will not occur again for many years. Therefore, space-based observation with observatories such as \Spitzer will be necessary to maintain the transit ephemeris. 

\begin{deluxetable}{cccc}
\tablecaption{Timing of Past and Future Transits of \textit{Kepler}-421b\label{tab:transits}} 
\tablehead{ 
\colhead{Number} & \colhead{$t_0$} & \colhead{$\sigma_{t_0}$} & \colhead{Potential} \\
\colhead{ }& \colhead{[UTC]} & \colhead{[hours]} & \colhead{Observatories\tablenotemark{a}} }\\ 
\startdata 
1 & 2012-04-12 11:10 & 0.05 & K \\
2 & 2014-03-17 15:56 & 0.05 & K \\
3 & 2016-02-19 20:42 & 0.34 & DCT \\
4 & 2018-01-24 01:27 & 0.67 & H, S \\
5 & 2019-12-29 06:13 & 1.01 & H, J \\
6 & 2021-12-02 10:59 & 1.35 & H, J \\
7 & 2023-11-06 15:44 & 1.68 & H, J \\
8 & 2025-10-10 20:30 & 2.02 & H, J \\
9 & 2027-09-15 01:16 & 2.36 & H, J 
\enddata 
\tablenotetext{a}{The potential observatories are \textbf{K}:\textit{Kepler Space Telescope}, \textbf{DCT}:Discovery Channel Telescope, \textbf{H}:\textit{Hubble Space Telescope}, \textbf{S}:\textit{Spitzer Space Telescope}, and \textbf{J}:\textit{James Webb Space Telescope}. We adopt optimistic estimates for the lifetimes of both \textit{HST} and \textit{JWST}.}
\end{deluxetable}

The importance of observations specifically aimed at aiding the recovery of future exoplanet transits was mentioned recently by \citet{Beichman2016}. This practice is imperative for long-period exoplanets. In a sample of long-period transiting exoplanets identified by the Planet Hunters project \citep{Wang2015}, at least 50\% displayed TTVs ranging from 2 to 40 hours. Beyond this small sample of known long-period transiting exoplanets, it is possible that exoplanets discovered in the future will display even larger TTVs. For these exoplanets, missing even a single transit can result in $t_0$ uncertainties of several days. With that level of error, subsequent follow-up becomes impossible and the planet is essentially \emph{lost} barring another long-term, dedicated transit survey covering the same portion of sky. 

\textit{Kepler}-421b, which resides near its system's snow-line, is an excellent test case for theories involving planetary formation and evolution \citep[e.g.][]{Oberg2011}. Atmospheric characterization of this cold exoplanet may constrain its mass \citep[e.g.][]{deWit2013} or offer details about ongoing processes such as photochemistry \citep[e.g.][]{Dalba2015}. However, \textit{Kepler}-421's faintness (J = 12 mag) and \textit{Kepler}-421b's small transit depth ($\sim$0.3\%) make it a challenging target for transmission spectroscopy with the \textit{Hubble Space Telescope} (\textit{HST}). Even if its cold (T$_{eq}$ = 185 K) atmosphere contains gaseous methane, which has an absorption feature at 1.4 $\mu$m (within the sensitivity of the Wide Field Camera 3), it is unlikely that a high confidence detection could be obtained in a single transit. Maintaining a precise transit ephemeris into the era of the \textit{James Webb Space Telescope} (\textit{JWST}) is therefore vital for future attempts to study \textit{Kepler}-421b.

In the near future, the \textit{Transiting Exoplanet Survey Satellite} (\textit{TESS}) is expected to find around a half-dozen planets with radii of 6 to 22 R$_{\earth}$ and periods of several hundred days \citep{Sullivan2015}. Some of these planets may be candidates for follow-up observation, but the process of confirmation and characterization will require several years at least---perhaps longer than the guaranteed five-year lifetime of \textit{JWST}. In this way, \textit{Kepler}-421b offers one of the best opportunities to investigate the unknown parameter space of long-period giant exoplanets.

\acknowledgements

The authors thank the anonymous referee for a thoughtful critique of this work. The authors also thank Brian Taylor and the DCT telescope operators for their assistance with the observations, and Bryce Croll, Andrew Vanderburg, Phillip Phipps, and Mark Veyette for helpful conversations involving the strategy and analysis of the observations. 

This research made use of the Discovery Channel Telescope at Lowell Observatory, supported by Discovery Communications, Inc., Boston University, the University of Maryland, the University of Toledo, and Northern Arizona University. The Large Monolithic Imager was funded by the National Science Foundation via grant AST-1005313. 

This research made use of the NASA Exoplanet Archive, which is operated by the California Institute of Technology, under contract with the National Aeronautics and Space Administration under the Exoplanet Exploration Program.

This research made use of analytic transit models based on the formalism from Mandel \& Agol (2002) and adapted for \texttt{Python} by Laura Kreidberg. 

{\it Facilities:} \facility{Discovery Channel Telescope, Large Monolithic Imager}

\end{document}